\documentclass[sigconf]{acmart}

\usepackage{enumitem}
\usepackage{tabularx}
\usepackage{graphicx}
\usepackage{amsmath}
\usepackage{multirow, multicol, makecell}
\usepackage{listings}
\usepackage[table]{xcolor}
\usepackage{framed}
\usepackage{hyperref}

\definecolor{baselinecolor}{RGB}{235,245,255}
\definecolor{cddcolor}{RGB}{235,255,235}

\copyrightyear{2026}
\acmYear{2026}
\acmConference[EASE 2026]{The 30th International Conference on Evaluation and Assessment in Software Engineering}{9–12 June, 2026}{Glasgow, Scotland, United Kingdom}
\acmDOI{10.1145/XXXXXXX.XXXXXXX}
\acmISBN{979-X-XXXX-XXXX-X/2026/XX}


\author{Subarna Saha}
\orcid{0009-0005-2414-8063}
\affiliation{%
  \institution{Jahangirnagar University}
  \city{Dhaka}
  \country{Bangladesh}
}
\email{subarna.stu2019@juniv.edu}

\author{Alif Al Hasan}
\orcid{0009-0000-3752-616X}
\affiliation{%
  \institution{Case Western Reserve University}
  \city{Cleveland, Ohio}
  \country{USA}
}
\email{alifal.hasan@case.edu}

\author{Fariha Tanjim Shifat}
\orcid{0009-0001-9305-5961}
\affiliation{%
  \institution{Missouri University of Science and Technology}
  \city{Rolla, Missouri}
  \country{USA}
}
\email{fsvfh@mst.edu}

\author{Mia Mohammad Imran}
\orcid{0000-0003-2477-9971}
\affiliation{%
  \institution{Missouri University of Science and Technology}
  \city{Rolla, Missouri}
  \country{USA}
}
\email{imranm@mst.edu}

\title[Improving Code Comprehension through Cognitive-Load Aware Automated Refactoring]{Improving Code Comprehension through Cognitive-Load Aware Automated Refactoring for Novice Programmers}

\begin{document}

\begin{abstract}
Novice programmers often struggle to comprehend code due to vague naming, deep nesting, and poor structural organization. While explanations may offer partial support, they typically do not restructure the code itself. We propose \textit{code refactoring as cognitive scaffolding}, where cognitively guided refactoring automatically restructures code to improve clarity. We operationalize this in \textsc{CDDRefactorER}, an automated approach grounded in \textit{Cognitive-Driven Development} that constrains transformations to reduce control-flow complexity while preserving behavior and structural similarity.

We evaluate \textsc{CDDRefactorER} using two benchmark datasets (MBPP and APPS) against two models (\texttt{gpt-5-nano} and \texttt{kimi-k2}), and a controlled human-subject study with novice programmers. Across datasets and models, \textsc{CDDRefactorER} reduces refactoring failures by 54-71\% and substantially lowers the likelihood of increased Cyclomatic and Cognitive complexity during refactoring, compared to unconstrained prompting. Results from the human study show consistent improvements in novice code comprehension, with function identification increasing by 31.3\% and structural readability by 22.0\%. The findings suggest that cognitively guided refactoring offers a practical and effective mechanism for enhancing novice code comprehension.
\end{abstract}

\maketitle

\section{Introduction}

Program comprehension is a central activity in software development and a persistent challenge for novice programmers~\cite{bois2005does, costa2023evaluating, johnson2019empirical, park2024eye, siegmund2017measuring}. 
Despite acquiring foundational syntactic and semantic knowledge, novice programmers frequently struggle to comprehend existing code, including identifying program purpose, tracing control-flow, and recognizing functional decomposition. Critically, prior research does not attribute these difficulties to syntactic unfamiliarity. It attributes them to structure, specifically, to deep nesting, complex control-flow, and unclear modular boundaries~\cite{busjahn2011analysis, sellitto2022toward, siegmund2017measuring, peitek2021program, fakhoury2018effect}. 

\begin{figure}[tb]
\centering
\includegraphics[width=\linewidth]{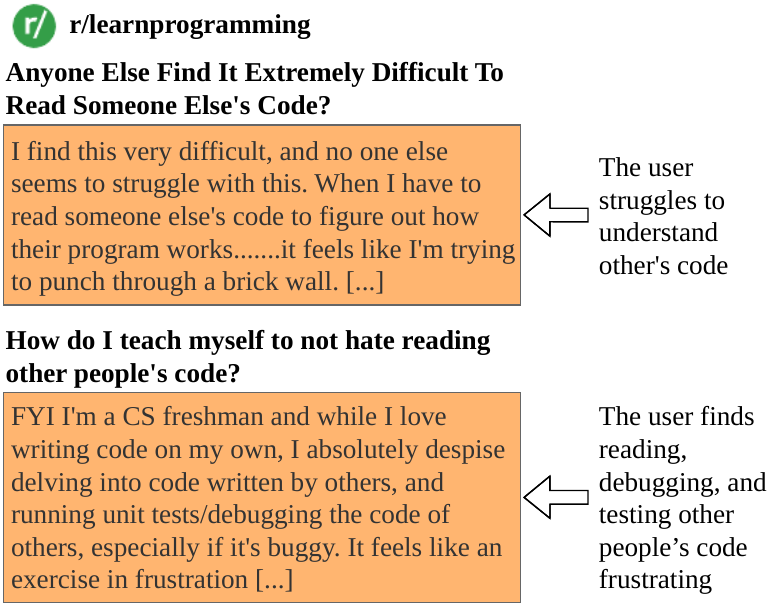}
\caption{Examples from Reddit where novice programmers talking about difficulties in understanding other's code}
\label{fig:motivating-example}
\end{figure}

Figure~\ref{fig:motivating-example} illustrates this pattern through examples drawn from public programming forums, where novice programmers consistently report that structural organization--not language syntax--is what defeats their attempts at comprehension. Recent empirical work corroborates this: structural breakdowns are a documented trigger for confusion and frustration among novices, and are closely associated with cognitive overload during learning activities~\cite{Hasan2026LearningProgramming}.

The structural account of novice difficulty motivates a structural intervention.
Refactoring is commonly used to improve code readability by restructuring code while preserving behavior~\cite{fowler1999improving}. 
If structural characteristics constitute the primary barrier to novice comprehension, then targeted restructuring should, in principle, lower that barrier. Prior empirical work, however, challenges this inference. Refactoring does not consistently improve novice comprehension~\cite{sellitto2022toward}. 
Approaches that emphasize structural reorganization or metric reduction without accounting for the cognitive effort required during comprehension~\cite{sellitto2022toward, morales2020repor, peruma2022refactor} may produce transformations that reduce conventional complexity metrics while simultaneously increasing the reasoning burden on novices--disrupting the control-flow paths and data dependencies they had begun to trace~\cite{carneiro2024investigating, wiese2019linking}. The efficacy of refactoring therefore depends not solely on behavioral preservation but on how structural modifications are constrained.

Cognitive-Driven Development (CDD) provides a principled basis for such constraints~\cite{souza2020toward}. CDD bounds control-flow complexity within individual code units according to working memory capacity limits~\cite{miller1956magical}, aligning program structure with the cognitive demands of comprehension. Prior work confirms that CDD constraints improve code readability and support developer reasoning~\cite{pinto2021cognitive, pinto2023cognitive, ferreira2024assisting}. However, manual application of CDD remains inconsistent in practice, and novice programmers cannot apply these constraints reliably without external scaffolding~\cite{techapalokul2019position, carneiro2024investigating}.

This gap motivates \textsc{CDDRefactorER}, an automated refactoring system that encodes CDD-inspired constraints into the prompting strategy of large language models. Unconstrained prompting produces inconsistent and sometimes counterproductive transformations. \textsc{CDDRefactorER}, by contrast, directs the model to identify code units that exceed cognitive thresholds and to apply refactoring strategies grounded in novice comprehension research, including method extraction, nesting reduction, identifier improvement, and sequential flow organization~\cite{adler2021improving, fakhoury2019improving, nurollahian2024teaching, carneiro2024investigating}, while preserving behavioral fidelity and structural similarity to the original code~\cite{hermans2016code, wiese2019linking, fowler1999improving}. \textsc{CDDRefactorER} does not perform semantic transformation~\cite{sciencedirect_code_transformation}, bug fixing, or program repair~\cite{le2019automated}.

To what extent, cognitive constraints improve refactoring safety, structural control, and novice comprehension requires empirical validation. We pursue that validation through the following three research questions:

\noindent\textbf{RQ1: Evaluation of Baseline Unconstrained Prompt.}
\textit{How does an unconstrained prompt perform in refactoring tasks intended to be novice-programmer friendly?}
We find that unconstrained prompting preserves functional correctness in most cases on novice-oriented benchmarks, but still produces non-trivial refactoring failures. An error analysis reveals that failures commonly arise from unintended logic alterations, injected domain assumptions, and small value discrepancies.

\noindent\textbf{RQ2: Validation of \textsc{CDDRefactorER}.}
\textit{How does \textsc{CDDRefactorER}-guided refactoring differ from unconstrained prompting in correctness and code structure?} Across two benchmark datasets and two language models, CDDRefactorER reduces refactoring failures by 54.40 to 71.23 percent relative to unconstrained prompting. It substantially lowers the likelihood of increases in cyclomatic and cognitive complexity during refactoring, and preserves higher structural similarity to the original code, indicating more controlled and stable transformations.

\noindent\textbf{RQ3: Impact on Comprehension.}
\textit{How does systematic automatic refactoring using \textsc{CDDRefactorER} affect novice programmers’ ability to understand code?}
Results from a controlled between-subject human study with 20 novice programmers show consistent improvements in self-reported code comprehension after interacting with \textsc{CDDRefactorER}. The largest gains are observed in function identification (+31.3\%) and structural readability (+22.0\%), while challenges related to unfamiliar programming concepts persist.

\noindent\textbf{Contributions.}
This paper makes two primary contributions: (i) it introduces \textsc{CDDRefactorER}, a cognitively constrained automated refactoring approach, and (ii) it provides empirical evidence of its effects on refactoring correctness, code structure, and novice code comprehension. 
\color{blue}
The replication package is publicly available~\cite{anon_zenodo_17201118}.
\color{black}

\section{Background and Related Work}

This section provides the theoretical and empirical context for our work. We first give an overview of Cognitive-Driven Development (CDD), which forms the basis of our refactoring constraints. We then review literature on cognitive load in programming with an emphasis on novice comprehension, and examine prior work on refactoring for code comprehension.

\subsubsection*{\textbf{Cognitive-Driven Development (CDD)}} \label{cdd-refactoring}
CDD~\cite{souza2020toward}, grounded in Cognitive Load Theory~\cite{sweller1988cognitive} and cognitive complexity research~\cite{campbell2018cognitive}, is a software development approach that constrains code structure based on limits of human working memory~\cite{miller1956magical}. 
Rather than optimizing for abstract structural metrics alone, CDD emphasizes bounding the cognitive effort required to reason about control-flow and nesting within individual code units~\cite{campbell2018cognitive}.
CDD quantifies structural complexity through Intrinsic Complexity Points (ICPs), which assign costs to control-flow constructs such as conditionals, loops, and their nesting depth. ICPs are computed by aggregating the contributions of control-flow constructs within a function. Each construct contributes a base cost, and additional cost is incurred through nesting. The resulting ICP value is compared against predefined thresholds to determine whether a code unit exceeds acceptable structural complexity~\cite{souza2020toward}. The following example illustrates how ICPs are assigned. Consider a function that checks whether a number is prime:

\begin{lstlisting}
def is_prime(n):
    if n <= 1: # +1 ICP (branch)
        return False
    else: # +1 ICP (branch)
        i = 2
        while i < n: # +1 ICP (loop)
            if n % i == 0:  # +1 ICP (branch inside loop)
                return False
            else: # +1 ICP (nested branch)
                i += 1
        return True
\end{lstlisting}

In this example, conditional branches and loops contribute to a total of five ICPs. The same logic can be expressed with fewer control-flow constructs:

\begin{lstlisting}
def is_prime(n):
    if n <= 1: # +1 ICP (branch)
        return False
    while i < n: # +1 ICP (loop)
        if n % i == 0: # +1 ICP (branch inside loop)
            return False
        i += 1
    return True
\end{lstlisting}

This version totals three ICPs due to the removal of nested branches, while preserving the original program behavior. CDD defines structural complexity using ICP counts and predefined thresholds.
Prior work shows that CDD benefits both professional software development tasks and novice programmers (e.g., picking up a new language)~\cite{ferreira2024assisting}. Researchers reported that manually applying CDD is helpful to improve code readability~\cite{barbosa2022to} and refactoring~\cite{pinto2021cognitive}.

\subsubsection*{\textbf{Cognitive Load in Programming}}  
Programming requires developers to reason about control-flow, data dependencies, and intermediate program state, which places demands on working memory~\cite{white2002theory}. 
Cognitive Load Theory distinguishes between intrinsic load, extraneous load, and germane load, and has been used to analyze programming tasks and learning outcomes~\cite{sweller1988cognitive}. Prior literature reviews highlight the prevalence of CLT in computing education research and emphasize strategies such as scaffolding to manage high element interactivity in code~\cite{berssanette2021cognitive, duran2022cognitive}. 

Neuroimaging and physiological studies demonstrate that code comprehension activates brain regions associated with working memory and attention, with increasing complexity correlating with higher neural load~\cite{fakhoury2018effect, fakhoury2019improving, roy2020model, gonccales2021measuring}. 
Emerging research on AI-assisted tools, such as GitHub Copilot, suggests that these systems can reduce cognitive load by automating repetitive coding tasks, allowing developers to focus on higher-level reasoning~\cite{barke2023grounded, ziegler2024measuring}. 
However, Prather et al. show that while such tools reduce syntactic burden for novices, they introduce additional metacognitive demands related to verification and understanding of generated code, indicating a shift rather than a net reduction in cognitive load~\cite{prather2023weird}. Recent studies suggest that when cognitive load remains unmanaged, novices experience persistent confusion and frustration, linking working memory overload to observable affective struggle~\cite{Hasan2026LearningProgramming}.

\subsubsection*{\textbf{Refactoring for Code Comprehension}}
Refactoring restructures source code to improve clarity and maintainability without altering external behavior~\cite{fowler1999improving}. 
Prior studies associate structural transformations such as decomposition, improved identifiers, and reduced cyclomatic complexity with lower cognitive load and improved program comprehension~\cite{sellitto2022toward, scalabrino2016improving, busjahn2011analysis, siegmund2017measuring}. 
Additional work examines finer-grained structural practices, including selective annotation and localized restructuring, with reported benefits for readability~\cite{gopstein2017understanding, medeiros2018discipline, schulze2013does}. 
However, empirical evidence shows that readability improvements do not consistently translate to better novice comprehension when refactoring disrupts structural familiarity or existing mental models~\cite{wiese2019linking}. 
Code smells are also shown to hinder novice performance, while constrained refactorings can improve learning outcomes~\cite{adler2021improving, hermans2016code}.

Manual refactoring remains difficult for novices and is associated with semantic errors~\cite{techapalokul2019position, wiese2019linking}. 
Think-aloud and replication studies report that novices reason locally and struggle to apply refactoring strategies without guidance~\cite{carneiro2024investigating, bennett2025replicating}. 
As a result, prior work has explored external scaffolding, including LLM-generated explanations, which support understanding while leaving code structure unchanged~\cite{feng2020codebert, chen2021evaluating, roziere2023code, macneil2023experiences}. 
More recent work investigates guided and LLM-based refactoring, reporting improved refactoring quality alongside sensitivity to prompting and novice over-trust in generated outputs~\cite{piao2025refactoring, palit2025reinforcement, ericsson2023evaluating, xu2025mantra, alomar2025chatgpt, carneiro2025uncovering}.

\smallskip
Although prior research has examined cognitive load in programming and automated refactoring independently, their integration for novice code comprehension remains underexplored, a gap this study aims to address.

\section{Methodology}\label{dataset-human-study}

\begin{figure*}
    \centering
    \includegraphics[width=0.85\linewidth]{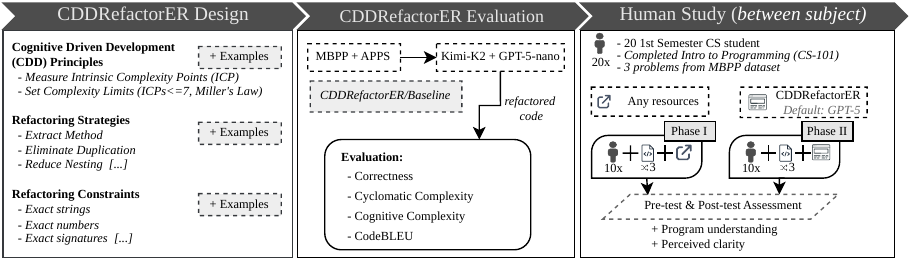}
    \caption{Overview of the Methodology.}
    \label{fig:methodology}
\end{figure*}

Figure~\ref{fig:methodology} shows the overview of the methodology. It has two components: (i) the design of \textsc{CDDRefactorER} and (ii) its evaluation. 

We design \textsc{CDDRefactorER} by incorporating CDD principles, as described in Section~\ref{cdd-refactoring}. 
We then evaluate the approach using two complementary studies. First, we benchmark two prompting strategies on the MBPP~\cite{austin2021program} and APPS~\cite{hendrycks2021apps} datasets: (i) an unguided zero-shot baseline prompt without structural constraints, and (ii) a CDD-guided prompt--\textsc{CDDRefactorER}. We evaluate these refactorings in terms of functional correctness and structural complexity, measured using cyclomatic and cognitive complexity metrics in \textbf{RQ1} and \textbf{RQ2}.

Second, we conduct a controlled human-subject study to assess the impact of CDD-guided refactoring on novice code comprehension (\textbf{RQ3}). The following sections describe the prompt design used in the empirical study and the human study methodology.

\subsection{Prompt Engineering}
We compare two prompting strategies for automated refactoring.

\subsubsection{\textbf{Unconstrained Zero-shot Prompt (Baseline)}}
The baseline prompt instructs the model to refactor code for readability and maintainability without imposing any explicit structural or cognitive constraints. It serves as a representative unconstrained refactoring approach. The prompt is included in the replication package~\cite{anon_zenodo_17201118}.

\begin{promptbox}{Baseline Prompt}
{
\noindent
You are an AI assistant specialized in refactoring code for novice programmers. Your goal is to make the code more readable, understandable, and maintainable. [...]
}
\end{promptbox}

\subsubsection{\textbf{\textsc{CDDRefactorER} Prompt}}\label{cdd-prompt-design}
The \textsc{CDDRefactorER} prompt operationalizes three CDD principles: defining Intrinsic Complexity Points (ICPs), constraining code complexity to human cognitive capacity, and reducing ICPs through refactoring. Following the original formulation, the prompt assigns ICP values~\cite{souza2020toward}. We assign the ICP values and ICP limits in a code block based on the work by de Souza et al.~\cite{souza2020toward} and their following works~\cite{pinto2021cognitive, pinto2022effects, pinto2023cognitive}.
The model is instructed to identify code units whose accumulated ICPs exceed acceptable thresholds as per Miller's law~\cite{miller1956magical} and to target these units for refactoring. This emphasis on control-flow is motivated by prior empirical evidence showing that nested conditionals and loops are particularly challenging for novice programmers to understand~\cite{siegmund2017measuring, sellitto2022toward, peitek2021program, fakhoury2019improving, fakhoury2018effect}.
The prompt further specifies a set of refactoring strategies grounded in prior research on code comprehension for novices. Extract Method decomposes complex functions into smaller, single-purpose units~\cite{hermans2016code, scalabrino2016improving}, while Reduce Nesting flattens deeply nested control structures~\cite{sellitto2022toward, wiese2019linking}. Eliminate Duplication factors out repeated code fragments~\cite{hermans2016code, wiese2019linking}, and Simplify Boolean Returns replaces verbose conditional patterns with direct boolean expressions~\cite{wiese2019linking}. Descriptive Naming improves identifier clarity~\cite{scalabrino2016improving, sellitto2022toward}, and Sequential Flow encourages chronological ordering and grouping of statements to support comprehension~\cite{sweller1988cognitive}. Each strategy is defined through explicit transformation rules and illustrated with concrete examples in the prompt.

Finally, we incorporated additional constraints derived from an error analysis of baseline prompt outputs (see: Section~\ref{sec:error-analysis}). These constraints are intended to prevent unintended semantic or stylistic alterations introduced by the model. In particular, generative outputs occasionally substitute domain-specific constants, such as replacing literal values like 3.14 with $\pi$, or modify string literals by altering capitalization, for example transforming `fizzbuzz' into `FizzBuzz'. The prompt explicitly discourages such changes to preserve functional behavior.

The full version of the CDDRefactorER prompt is provided in the replication package~\cite{anon_zenodo_17201118}, and a shortened version is shown below.


\begin{promptbox}{\textsc{CDDRefactorER} Prompt (Short Version)}
{
\noindent
You are CDDRefactorER, an AI that refactors code to reduce cognitive load for novice programmers while preserving exact behavior.

\medskip
\noindent
\textbf{CDD Principles.}
\begin{enumerate}[leftmargin=*]
  \item \textbf{Measure ICPs:} Control structures (\texttt{if}: +1; [...]
  \item \textbf{Set Complexity Limits:} Keep ICPs $\leq$ 7 per function (Miller's Law: humans hold ~7±2 items in working memory). [...]
  \item \textbf{Refactor When Exceeded:} Decompose complex units into simpler, focused functions.  [...]
\end{enumerate}

\smallskip
\noindent
\color{blue} Example [...] \color{black}

\medskip
\noindent
\textbf{Refactoring Strategies.}
\begin{itemize}[leftmargin=*]
  \item \textbf{Extract Method:} Break complex functions into single-purpose helpers.
  \item \textbf{Eliminate Duplication:} Factor out repeated code. Example: \texttt{setup(); (a() if x else b()); cleanup()}
  \item \textbf{Improve Naming:} Use \texttt{verb\_noun} for functions, \texttt{is/has/can} for booleans.
  \item \textbf{Reduce Nesting:} [...] 
\end{itemize}
\color{blue} Examples [...] \color{black}

\medskip
\noindent
\textbf{Constraints (Do not violate):} [...]
\begin{itemize}[leftmargin=*]
  \item \textbf{Exact strings:} ``fizzbuzz'' must not become ``Fizzbuzz'' or ``FizzBuzz''.
  \item \textbf{Exact numbers:} 3.14 must not become \texttt{math.pi}.
  \item \textbf{Exact signatures:} Don't change function names, parameters, or order. [...]
\end{itemize}

\noindent
\color{blue} Examples [...] \color{black}

\medskip
\noindent
\textbf{Task:} Now for the given code snippet, do Code Refactoring using above guideline.
[...]
}
\end{promptbox}

\subsubsection{\textbf{Model}} We evaluated our strategy using one proprietary model and one open-source model:

\begin{itemize}[leftmargin=*]
    \item \textbf{gpt-5-nano.} We use the \texttt{gpt-5-nano} model to evaluate \textsc{CDDRefactorER}. This model was released on August 7, 2025. 
    \item \textbf{kimi-k2.} \texttt{kimi-k2} is an open-source Mixture-of-Experts language model with 32 billion active parameters drawn from a larger expert pool. It reported strong performance on competitive coding benchmarks~\cite{team2025kimi}.
\end{itemize}

Both models are evaluated using identical prompts and experimental conditions to isolate the effect of prompting strategy.
 
\subsubsection{\textbf{Dataset}} We evaluated our approach against two datasets:

\begin{itemize}[leftmargin=*]
    \item \textbf{MBPP dataset.} 
The MBPP dataset contains 974 introductory-level Python programs, each paired with a problem description, reference implementation, and test cases. The dataset primarily targets novice-level programming tasks and is well suited for evaluating refactoring correctness and structural changes~\cite{austin2021program}.
\item \textbf{APPS Dataset}. 
The APPS dataset is a large-scale benchmark consisting of 10,000 programming problems and over 230k human-written solutions~\cite{hendrycks2021apps}. We restrict our analysis to the introductory subset of APPS and randomly sample 5,000 solutions from this subset to ensure tractability while maintaining diversity.
\end{itemize}

\subsubsection{\textbf{Metrics}} We assess refactoring outcomes using multiple complementary metrics.

\noindent \textbf{Functional correctness.} Correctness is measured by running the refactored programs against the original test suites provided with each dataset. A refactored program is considered correct only if it passes all associated test cases.

\noindent
\textbf{Cyclomatic Complexity (CC).}
Cyclomatic complexity measures the number of independent control-flow paths in a program~\cite{mccabe1976complexity}. It is defined on the control-flow graph $G$ as $V(G) = E - N + 2P$, where $E$ is the number of edges, $N$ is the number of nodes, and $P$ is the number of connected components. 
We use this metric to capture changes in control-flow structure between the original and refactored code.

\noindent
\textbf{Cognitive Complexity (CogC).} 
Cognitive complexity captures how difficult a program’s control-flow is to understand by accounting for control constructs and their nesting depth~\cite{campbell2018cognitive}. 
It is defined as $\text{CogC} = \sum_{i=1}^{n} (1 + d_i)$, where $n$ is the number of control-flow structures (e.g., \texttt{if}, \texttt{for}, \texttt{while}, \texttt{catch}), and $d_i$ represents the nesting depth of the $i$-th structure.
We use this metric to assess how refactoring affects nesting and control-flow complexity relative to the original code.

\noindent
\textbf{Statistical Significance ($p$).}
To test whether differences between baseline and refactored code are statistically significant, we use non-parametric two-sided Wilcoxon Signed-Rank Test~\cite{wilcoxon1945individual}.

\noindent
\textbf{Effect Size.}
Effect size measures the magnitude of differences between conditions. We report Cliff’s Delta ($\delta$)~\cite{cliff1993dominance}, a non-parametric effect size measure for comparing two distributions. 
$\delta$ values are interpreted as negligible ($|\delta| < 0.147$), small ($0.147 \le |\delta| < 0.33$), medium ($0.33 \le |\delta| < 0.474$), or large ($|\delta| \ge 0.474$)~\cite{romano2006appropriate}.

\noindent
\textbf{CodeBLEU.} 
We measure syntactic and semantic similarity between the original and refactored code using CodeBLEU~\cite{ren2020codebleu}. CodeBLEU extends BLEU by incorporating code-specific features such as $n$-gram overlap, Abstract Syntax Tree (AST) structure, and data-flow information. 
CodeBLEU produces a weighted similarity score. Higher scores indicate greater structural and syntactic similarity between the two programs.

\subsection{Human Study Design}\label{sec:human-study}

The goal of the human study is to assess how cognitively guided refactoring influences novice programmers’ code comprehension. We focus on understanding program purpose, logic flow, functional decomposition, and structural readability.

\subsubsection{\textbf{Participants}}
We recruited 20 first-semester computer science students (6 male, 14 female) who had completed an introductory programming course (CS-101) and had no advanced coursework. Participants had between 0 and 2 years of programming experience and represent novice programmers with foundational but limited exposure. Participation was voluntary, and informed consent was obtained under Institutional Review Board (IRB) approval.

\subsubsection{\textbf{Study Design}}
The study employed a between-subjects design consisting of two independent groups: a pre-test group and a post-test group~\cite{betweenvswithin}. Each group included 10 participants, and no individual participated in both conditions. This design was selected to avoid learning, familiarity, and testing effects that may arise from repeated exposure to the same code artifacts or survey instruments~\cite{charness2012experimental, greenwald1976within}.

We informed the participants that the study evaluated code comprehension rather than code writing performance, debugging ability, or task completion speed.

\subsubsection{\textbf{Task Sampling and Allocation}}
\noindent \textbf{Problem Selection.} 
Tasks were drawn from the MBPP dataset using a multi-stage selection process. Three authors independently selected candidate problems spanning three difficulty levels:

\begin{itemize}[leftmargin=*]
    \item \textbf{Basic tasks} required simple arithmetic or string manipulation with minimal control-flow.
    \item \textbf{Intermediate tasks} involved standard data structures such as lists or dictionaries, or nested loops.
    \item \textbf{Advanced tasks} required non-trivial algorithmic reasoning or careful edge-case handling.
\end{itemize}
After merging selections, we obtained a pool of 81 candidate problems. The authors held discussion sessions and reached consensus on task difficulty classifications based on algorithmic structure and required prior knowledge. After reaching unanimous agreement, we randomly selected a final set of 20 tasks consisting of 10 basic, 5 intermediate, and 5 advanced tasks.

We assigned each participant three tasks: one basic, one intermediate, and one advanced. We randomized task assignments while ensuring comparable difficulty distributions across the two experimental groups. We set an upper bound of 15 minutes per task.

\subsubsection{\textbf{Platform and Tooling}}
We delivered all tasks and surveys through a custom Streamlit web application that controlled task presentation, resource access, and response collection. We deployed \textsc{CDDRefactorER} through the OpenAI platform using the default \texttt{gpt-5} model~\cite{cddrefactorgpt2025}.

\subsubsection{\textbf{Procedure}}
The experiment followed a two-phase between-subjects procedure~\cite{betweenvswithin}:  

\noindent
\textbf{Phase 1:} The 10 randomly selected participants analyzed original, unrefactored code snippets. Their task was to understand the program’s purpose, logic flow, and structure. To reflect realistic novice learning behavior, participants were permitted to consult external online resources, including general web search engines, ChatGPT and other AI tools, and programming-related websites. 

\noindent
\textbf{Phase 2:}
The other 10 participants first examined the original, unrefactored code snippets. They then used \textsc{CDDRefactorER} to generate a refactored version of the code. After generating the refactored code, participants analyzed both the original and the refactored versions to understand the program's purpose, logic flow, and structure. During this phase, the participants restricted themselves to use \textsc{CDDRefactorER} only. Before beginning, they viewed a short instructional video explaining the purpose of \textsc{CDDRefactorER} and how to use the tool.

\subsubsection{\textbf{Surveys}}
Both pre-test and post-test conditions included code comprehension assessments administered after each task.

\noindent
\textbf{\textit{Pre-test surveys consisted of:}}
\begin{itemize}[leftmargin=*]
    \item Open-ended questions asking participants to describe what the code does and identify confusing sections.
    \item A 5-point Likert scale (ranging from Very Low - 1 to Very High - 5) measuring:
    \begin{itemize}[leftmargin=*]
        \item Perceived problem difficulty.
        \item Understanding of overall program purpose.
        \item Understanding of program logic flow.
        \item Ability to identify key functions and their roles.
        \item Perceived structural clarity and readability.
    \end{itemize}
\end{itemize}

\noindent
\textbf{\textit{Post-test surveys:}} repeated the Likert-scale comprehension questions after participants reviewed the refactored code. Open-ended questions asked participants to describe the refactored code and identify sections that became clearer after using \textsc{CDDRefactorER}.

Each participant completed surveys for three tasks, yielding 30 pre-test and 30 post-test responses. All collected data were anonymized to protect participant privacy.

\section{Evaluation of Baseline}

\textbf{RQ1:} \textbf{\textit{How does an unconstrained prompt perform in refactoring tasks intended to be novice-programmer friendly?}}

To evaluate the performance of an unguided refactoring prompt on novice-oriented tasks, we first examine functional correctness, defined as whether the refactored program preserves the original behavior as validated by the MBPP test suite. Using a simple unconstrained zero-shot refactoring prompt with the \texttt{gpt-5} model, we apply this baseline setting to 974 programs from the MBPP dataset. Under this criterion, the baseline successfully refactors 938 programs, corresponding to a high success rate and indicating that unconstrained prompting can often preserve functional correctness for short, well-scoped programs typical of novice benchmarks. However, functional incorrectness still occurs in 36 cases (3.70\%). We analyze these 36 failing cases further in detail.

In addition to correctness, we examine how unconstrained refactoring affects code structure (see Table~\ref{tab:complexity}). Across the refactored programs, cognitive complexity increases in 229 cases and decreases in 231 cases, resulting in a net change of $-2$. Similarly, cyclomatic complexity increases in 184 cases and decreases in 232 cases, yielding a net change of $-48$. These results suggest that, while unconstrained refactoring can simplify structure, it does not consistently reduce complexity and may introduce structural regressions in a non-trivial number of cases.


\begin{table}[tb]
\centering
\caption{Error categories observed in MBPP refactoring outputs generated by the \texttt{gpt-5} model on Baseline approach.}
\setlength{\tabcolsep}{1pt} 
\footnotesize
\begin{tabular}{|p{4cm}|c|}
\hline
Error Category & count \\
\hline
Logic alteration & 19 \\
Small value discrepancy & 7 \\
Function signature changes & 4 \\
Conditional logic issues & 2 \\
Miscellaneous & 4 \\
\hline
Total & 36\\
\hline
\end{tabular}
\label{tab:error-analysis}
\end{table}

\subsubsection*{\textbf{Error Analysis}}\label{sec:error-analysis}
To understand the sources of functional correctness failures, one author independently inspected the refactored programs and labeled observed errors using open coding, without relying on a predefined taxonomy, following established qualitative analysis practices~\cite{khandkar2009open}. A second author then reviewed the derived error categories and their assignments. Any disagreements were discussed until consensus was reached. 

Table~\ref{tab:error-analysis} summarizes the distribution of identified error categories. The most prevalent category, accounting for 50\% of baseline failures, is \textit{logic alteration}. These errors occur when the language model introduces incorrect logic, often based on ambiguous function naming or external domain knowledge. For example, a function named \texttt{`$avg(a, b)$'} that originally returns \texttt{`$a + b$'} may be ``corrected'' by the model to return \texttt{`$(a + b) / 2$'} due to its prior knowledge, causing the refactored program to fail the test cases despite preserving syntactic correctness.

The \textit{small value discrepancy} category captures errors resulting from changes in numeric constants (e.g., replacing an approximate value of $\pi$ with $math.pi$) or from precision drift due to reordering arithmetic operations. The \textit{function signature changes} category, which appears only under the baseline prompt (4/38 cases), reflects cases where the model incorrectly assumes input parameter types or modifies the function signature. \textit{Conditional logic issues} involve the introduction of additional input checks during refactoring, such as enforcing constraints on parameter values that were not present in the original implementation (e.g., requiring parameters to an \textit{average(a, b)} function to be positive). Finally, the \textit{miscellaneous} category includes a range of failures, such as syntax errors, parsing errors, and uninitialized variables.

\smallskip
\noindent\fbox{
\parbox{0.46\textwidth}{
        \textit{\underline{Summary of RQ1.}} The results show that while functional correctness is preserved in most cases, a non-trivial number of refactorings fail due to logic alterations, injected assumptions, and small value discrepancies. Structurally, complexity reductions and increases largely cancel out, resulting in little net simplification. Overall, unconstrained prompting does not reliably produce refactorings aligned with novice comprehension needs.
    }
}

\section{Evaluation of \textsc{CDDRefactorER}}
As discussed in Section~\ref{cdd-prompt-design}, the design of \textsc{CDDRefactorER} is informed by the error patterns observed under unconstrained refactoring as well as CDD principles. 
Assessing comprehension after automated refactoring for beginners requires more than verifying functional correctness, as refactoring may increase structural complexity, thereby hindering code comprehension even when behavior is preserved.
While reductions in code complexity do not guarantee improved understanding, prior work shows that these metrics are associated with increased comprehension effort and perceived mental difficulty~\cite{hao2023accuracy, esposito2025early, munoz2020empirical}. 
Further, extensive or disruptive structural changes may hinder comprehension by reducing familiarity with the original code structure~\cite{hermans2016code, wiese2019linking}.
Accordingly, we evaluate refactoring quality in terms of i) functional correctness (Table~\ref{tab:correctness}), ii) changes in structural complexity (Table~\ref{tab:complexity}), and iii) structural similarity (Figure~\ref{fig:codebleu-boxplots}).

\subsection{RQ2.A: Functional Correctness}
\noindent
\textbf{\textit{How does \textsc{CDDRefactorER}-guided refactoring differ from unconstrained prompting in terms of functional correctness?}}
\begin{table}[tb]
\centering
\caption{Comparison of Incorrect Refactorings Between \textsc{CDDRefactorER} and the Baseline (The gray-shaded row denotes the error analysis
from this baseline configuration used to inform the design of \textsc{CDDRefactorER}).}
\label{tab:correctness}
\setlength{\tabcolsep}{1pt} 
\footnotesize
\begin{tabular}{|l|c|cc|c|}
\hline
Dataset & Model & {\sc CDDRefactorER} & {Baseline} & Error Change \\ \cline{3-4}
($N$) & & \multicolumn{2}{c|}{(Incorrect Count)} & (Reduction Rate) \\
\hline
\rowcolor{lightgray} MBPP & \texttt{gpt-5-nano} & \textbf{9 (0.92\%)} & 36 (3.70\%) & $-$2.78\% (75.00\%) \\
(974) & \texttt{kimi-k2} & \textbf{11 (1.13\%)} & 39 (4.01\%) & \textbf{$-$2.87\%} (71.79\%) \\
\hline
APPS & \texttt{gpt-5-nano} & \textbf{83 (1.66\%)} & 182 (3.64\%) & $-$1.98\% (54.40\%) \\
(5000) & \texttt{kimi-k2} & \textbf{107 (2.14\%)} & 372 (7.44\%) & \textbf{$-$5.3\% (71.23\%)} \\
\hline
\end{tabular}
\end{table}

Table~\ref{tab:correctness} reports the number of refactored programs that fail their associated test suites on the MBPP and APPS datasets under baseline and \textsc{CDDRefactorER}. 
Although MBPP with \texttt{gpt-5} results are reported for completeness, 
we exclude this configuration from the discussion since error analysis from this configuration directly informed the design of \textsc{CDDRefactorER}.

Across all settings, \textsc{CDDRefactorER} consistently reduces the number of refactoring failures relative to the baseline. On MBPP using \texttt{kimi-k2}, \textsc{CDDRefactorER} produces 963 functionally correct refactorings and 11 failures, compared to 935 correct refactorings and 39 failures under the baseline prompt, corresponding to a 71.79\% reduction in errors. On the APPS dataset with \texttt{gpt-5}, \textsc{CDDRefactorER} yields 4,917 correct refactorings and 83 failures, compared to 4,818 correct refactorings and 182 failures under unconstrained prompting, resulting in a 54.40\% reduction. Similarly, for \texttt{kimi-k2} on APPS, the number of failing refactorings decreases from 372 to 107, corresponding to a 71.23\% reduction in errors.

\subsection{RQ2.B: Code Structural Change Analysis}
\textbf{\textit{How does \textsc{CDDRefactorER}-guided refactoring differ from unconstrained prompting in terms of code structure?}}

We measure complexity using cognitive and cyclomatic metrics. Increases reflect added structural complexity, while decreases indicate simplification.
Structural similarity is assessed using CodeBLEU.
It serves as a proxy for the extent of structural change. 
Higher CodeBLEU scores indicate closer adherence to the original program structure, while lower scores reflect more reorganization. 

\subsubsection{\textbf{Code Complexity Analysis}}
\begin{table}[tb]
\centering
\caption{Impact of Baseline and \textsc{CDDRefactorER} Refactoring on Cognitive and Cyclomatic Complexity
($NS$, *, **, ***, **** indicate $p$ $\geq$ 0.05, $p$ < 0.05, $p$ < 0.01, $p$ < 0.001, and $p$ < 0.0001, respectively.
${\circ}$, ${\dagger}$, ${\ddagger}$, and ${\S}$ indicates negligible, small, medium, and large effect sizes).
}
\label{tab:complexity}
\setlength{\tabcolsep}{1pt} 
\footnotesize
\begin{tabular}{|p{1cm}|p{1.3cm}|l|p{1.2cm}|l|l|}
\hline
Dataset & Model & Metric & Measure & Baseline & \textsc{CDDRefactorER} \\
\hline
\Xcline{1-6}{1pt}
\rowcolor{lightgray} \cellcolor{white} &  &  & Decrease  & 229 (23.51\%)        & 170 (17.45\%)             \\ \cline{4-6}
\rowcolor{lightgray} \cellcolor{white} &  &  & Increase  & 231 (23.72\%)        & 85 (8.73\%)               \\ \cline{4-6}
\rowcolor{lightgray} \cellcolor{white} &  &  & NET (\%)  & -2 (-0.21\%)         & 85 (8.73\%)               \\ \cline{4-6}
\rowcolor{lightgray} \cellcolor{white} &  &  & $p$-value & NS                   & **                        \\ \cline{4-6}
\rowcolor{lightgray} \cellcolor{white} &  & \multirow{-5}{5em}{Cognitive Complexity (CogC)}& Cliff's $\delta$   & 0.024 ${\circ}$      & 0.180 ${\dagger}$         \\ \Xcline{3-6}{1pt}

\rowcolor{lightgray} \cellcolor{white} &  &  &  Decrease & 184 (18.89\%)        & 193 (19.82\%) \\ \cline{4-6}
\rowcolor{lightgray} \cellcolor{white} &  & &  Increase  & 232 (23.82\%)        & 42 (4.31\%) \\ \cline{4-6}
\rowcolor{lightgray} \cellcolor{white} &  & & NET (\%)   & -48 (-4.93\%)        & 151 (15.50\%) \\ \cline{4-6}
\rowcolor{lightgray} \cellcolor{white} &  & & $p$-value  & **                   & **** \\ \cline{4-6}
\rowcolor{lightgray} \cellcolor{white} & \multirow{-10}{1cm}{\texttt{gpt-5-nano}} & \multirow{-5}{5em}{Cyclomatic Complexity (CC)} & Cliff's $\delta$   & 0.159 ${\dagger}$    & 0.613 ${\S}$ \\ 
\Xcline{2-6}{1pt}

 &  &  &  Decrease          & \textbf{223 (22.9\%)}& 217 (22.28\%)                   \\ \cline{4-6}
 &  &  &  Increase          & 139 (14.3\%)         & \textbf{43 (4.41\%)}                     \\ \cline{4-6}
 &  &  & NET (\%)           & 84 (8.62\%)          & \textbf{174 (17.86\%)} \cellcolor{green}         \\ \cline{4-6}
 &  &  & $p$-value          & ***                  & ****                            \\ \cline{4-6}
 &  & \multirow{-5}{5em}{CogC} 
     & Cliff's $\delta$     & 0.195 ${\dagger}$    & \textbf{0.539 ${\S}$}  \cellcolor{green}         \\ \Xcline{3-6}{1pt}

 &  &  &  Decrease          & 155 (15.9\%)         & \textbf{195 (20.02\%)}                          \\ \cline{4-6}
 &  &  &  Increase          & 119 (12.2\%)         & \textbf{13 (1.33\%)}                            \\ \cline{4-6}
 &  &  & NET (\%)           & 36 (3.70\%)          & \textbf{182 (18.69\%)}  \cellcolor{green}               \\ \cline{4-6}
 &  &  & $p$-value          & *                    & ****                                   \\ \cline{4-6}
\multirow{-20}{*}{MBPP} & \multirow{-10}{*}{\texttt{kimi-k2}} &\multirow{-5}{5em}{CC}& Cliff's $\delta$     
                            & 0.136 ${\circ}$      & \textbf{0.777 ${\S}$}    \cellcolor{green}              \\ \Xcline{1-6}{1pt}

\multirow{20}{*}{APPS} & \multirow{10}{1cm}{\texttt{gpt-5-nano}} & \multirow{5}{5em}{CogC} &  Decrease            
                            & \textbf{1746 (34.92\%)} & 1323 (26.46\%) \\ \cline{4-6}
 &  & &  Increase           & 1454 (29.08\%)          & \textbf{616 (12.32\%)} \\ \cline{4-6}
 &  & & NET (\%)            & 292 (5.84\%)            & \textbf{707 (14.14\%)} \cellcolor{green} \\ \cline{4-6}
 &  & & $p$-value           & **                      & **** \\ \cline{4-6}
 &  & &  Cliff's $\delta$   & 0.057 ${\circ}$         & \textbf{0.234 ${\dagger}$} \cellcolor{green} \\ 
\Xcline{3-6}{1pt}

 &  & \multirow{5}{5em}{CC} &  Decrease            
                            & \textbf{1272 (25.44\%)} & 1226 (24.52\%) \\ \cline{4-6}
 &  & &  Increase           & 1328 (26.56\%)          & \textbf{309 (6.18\%)} \\ \cline{4-6}
 &  & & NET (\%)            & -56 (-1.12\%)           & \textbf{917 (18.34\%)} \cellcolor{green} \\ \cline{4-6}
 &  & & $p$-value           & *                       & **** \\ \cline{4-6}
 &  & & Cliff's $\delta$    & 0.039 ${\circ}$         & \textbf{0.534 ${\S}$} \cellcolor{green} \\ 
\Xcline{2-6}{1pt}

 & \multirow{10}{*}{\texttt{kimi-k2}} & \multirow{5}{5em}{CogC}&  Decrease            
                            & \textbf{1889 (37.8\%)}  & 1571 (31.42\%)                                  \\ \cline{4-6}
 &  & &  Increase           & 1089 (21.8\%)           & \textbf{506 (10.12\%)}                          \\ \cline{4-6}
 &  & & NET (\%)            & 800 (16.00\%)           & \textbf{1065 (21.30\%)} \cellcolor{green}       \\ \cline{4-6}
 &  & & $p$-value           & ****                    & ****                                            \\ \cline{4-6}
 &  & &  Cliff's $\delta$   & 0.209 ${\dagger}$       & \textbf{0.391 ${\ddagger}$} \cellcolor{green}   \\ \Xcline{3-6}{1pt}

 &  & \multirow{5}{5em}{CC} &  Decrease            
                            & \textbf{1409 (28.2\%)}  & 1397 (27.94\%)                              \\ \cline{4-6}
 &  &  &  Increase          & 751 (15.0\%)            & \textbf{183 (3.66\%)}                       \\ \cline{4-6}
 &  &  & NET (\%)           & 658 (13.16\%)           & \textbf{1214 (24.28\%)} \cellcolor{green}   \\ \cline{4-6}
 &  &  &  $p$-value         & ****                    & ****                                        \\ \cline{4-6}
 &  &  &  Cliff's $\delta$  & 0.291 ${\dagger}$       & \textbf{0.685 ${\S}$} \cellcolor{green}      \\ 
\Xcline{1-6}{1pt}

\end{tabular}
\end{table}

\begin{figure}[t]
    \centering
    \includegraphics[width=\linewidth]{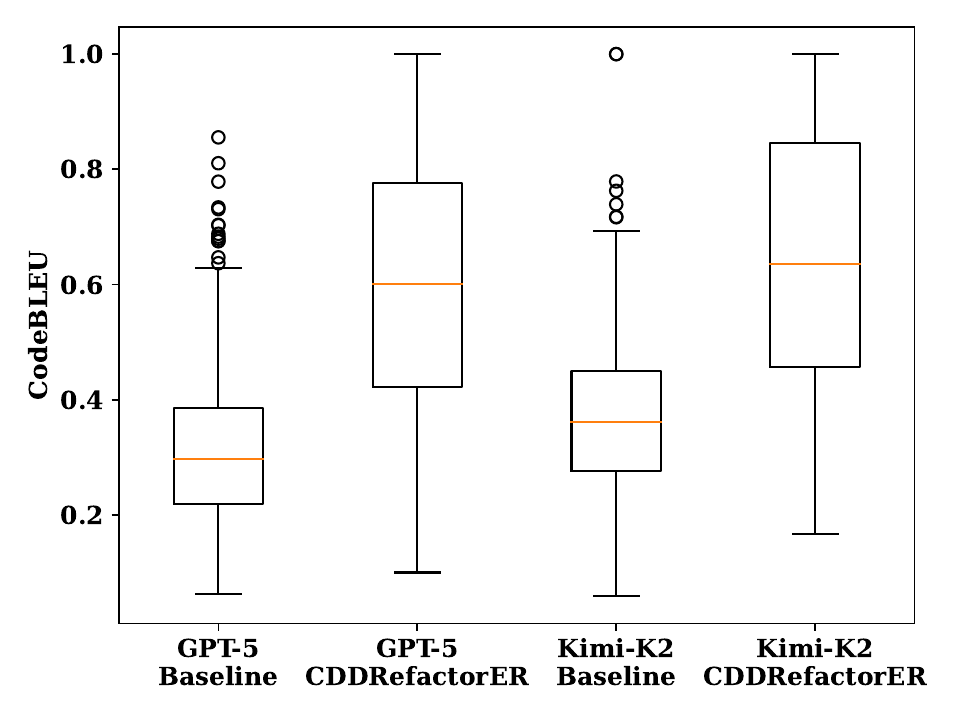}
    \includegraphics[width=\linewidth]{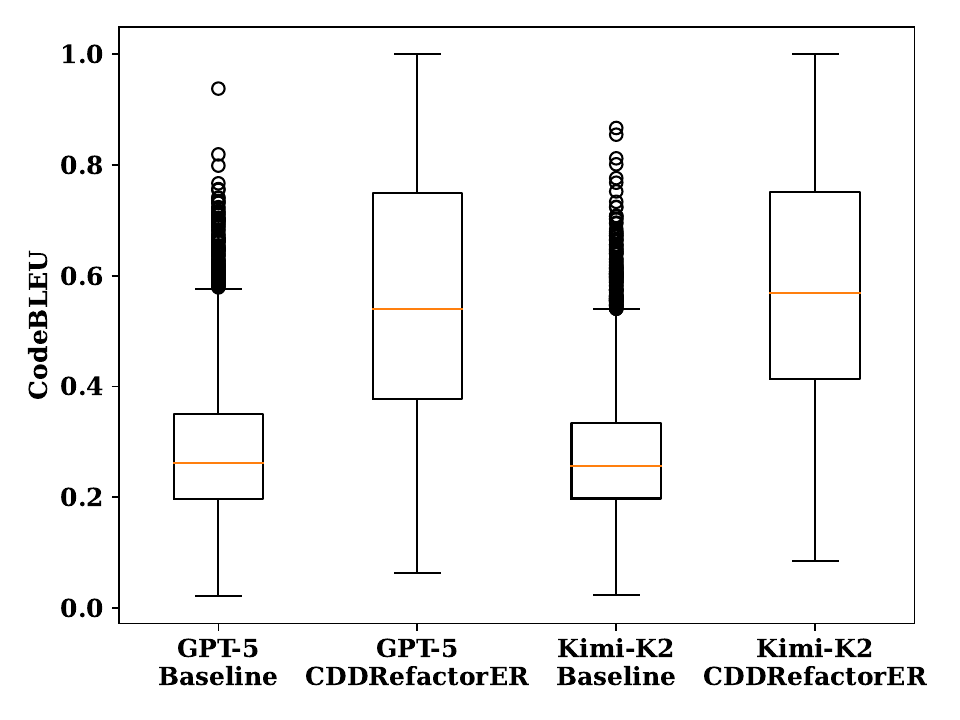}
    \caption{CodeBLEU similarity distributions after refactoring on MBPP (top) and APPS (bottom).}
    \label{fig:codebleu-boxplots}
\end{figure}

Table~\ref{tab:complexity} summarizes the impact of refactoring on cognitive and cyclomatic complexity under baseline prompting and \textsc{CDDRefactorER}. For each configuration, we report the proportion of refactorings that decrease or increase complexity, allowing us to assess whether refactoring tends to simplify or complicate program structure relative to the original implementation. The NET effect quantifies the balance between complexity-decreasing and complexity-increasing refactorings, expressed as the percentage difference between the two.

As with functional correctness, we include MBPP results for the \texttt{gpt-5} model in the table using \textsc{CDDRefactorER} for completeness but exclude from analysis.

\smallskip
\noindent
\textbf{Baseline behavior.}
Reductions and increases in complexity largely offset each other, resulting in limited net structural simplification. For example, on the APPS dataset with \texttt{gpt-5}, cognitive complexity decreases in $34.92\%$ of cases and increases in $29.08\%$ of cases, yielding a NET effect of $+5.84\%$. For cyclomatic complexity on the same dataset, decreases occur in $25.44\%$ of cases while increases occur in $26.56\%$, resulting in a negative NET effect of $-1.12\%$. Similar offsetting patterns are observed across datasets and models, indicating that unconstrained refactoring does not reliably prevent structural regressions. While most baseline configurations are statistically significant, all associated effect sizes are negligible to small, indicating limited separation between decreasing and increasing outcomes.

\smallskip
\noindent
\textbf{\textsc{CDDRefactorER} behavior.}
In contrast, \textsc{CDDRefactorER} consistently produces positive NET effects across datasets and models by substantially reducing the proportion of complexity-increasing refactorings. On the APPS dataset with \texttt{gpt-5}, cognitive complexity increases drop from $29.08\%$ under the baseline to $12.32\%$ with \textsc{CDDRefactorER}, while decreases occur in $26.46\%$ of cases, yielding a NET effect of $+14.14\%$. For cyclomatic complexity, increases are reduced from $26.56\%$ to $6.18\%$, while decreases remain comparable ($24.52\%$), resulting in a NET effect of $+18.34\%$. On APPS with \texttt{kimi-k2}, NET effects reach $+21.30\%$ for cognitive complexity and $+24.28\%$ for cyclomatic complexity. All \textsc{CDDRefactorER} configurations are statistically significant, with effect sizes ranging from medium to large (except one configuration which is small).

\subsubsection{\textbf{CodeBLEU Analysis}}
Figure~\ref{fig:codebleu-boxplots} presents CodeBLEU distributions for both datasets and models. Across all settings, \textsc{CDDRefactorER} consistently produces refactored code that remains closer to the original implementation than baseline refactoring.

In terms of central tendency, median CodeBLEU scores increase substantially under \textsc{CDDRefactorER} across both datasets and models. On MBPP dataset, the median rises from 0.297 to 0.601 for \texttt{gpt-5}, corresponding to a relative increase of $102.6\%$, and from $0.362$ to $0.635$ for \texttt{kimi-k2}, an increase of $75.6\%$. On APPS dataset, median similarity increases from 0.261 to 0.540 for \texttt{gpt-5}, a gain of $106.7\%$, and from $0.256$ to $0.570$ for \texttt{kimi-k2}, a gain of $122.7\%$. These results indicate that \textsc{CDDRefactorER} consistently preserves a substantially larger fraction of the original program structure during refactoring across both datasets and models.

Distributional analysis further supports this observation. On MBPP, the first quartile (Q1) increases from 0.219 to 0.421 for \texttt{gpt-5} and from 0.277 to 0.457 for \texttt{kimi-k2}, while the Q3 rises from 0.386 to 0.777 and from 0.449 to 0.846, respectively. On APPS, \texttt{gpt-5} exhibits an increase in Q1 from 0.197 to 0.377 and in Q3 from 0.350 to 0.750, while \texttt{kimi-k2} shows corresponding increases from 0.198 to 0.413 and from 0.335 to 0.750. 
The upward shift of both the lower and upper quartiles shows that \textsc{CDDRefactorER} yields higher structural similarity for most refactored programs.

\begin{figure}[tb]
\definecolor{boxcolor}{HTML}{CEBD99}
\colorlet{shadecolor}{boxcolor}
\begin{shaded}
\begin{minipage}[t]{\linewidth}
\textbf{Original Code:}
\begin{lstlisting}
def parallel_lines(line1, line2):
    return line1[0]/line1[1] == line2[0]/line2[1]            
\end{lstlisting}
\end{minipage}%
\hfill

\begin{minipage}[t]{\linewidth}
    \textbf{Baseline - Refactored using \texttt{gpt-5-nano}}:
\begin{lstlisting}
def parallel_lines(line1, line2):
    n1, d1 = line1
    n2, d2 = line2
    # Both lines have vertical slope ( undefined ): considered parallel
    if d1 == 0 and d2 == 0:
        return True
    # One vertical, the other not: not parallel
    if d1 == 0 or d2 == 0:
        return False
    # Compare slopes without floating point precision issues : n1/d1 == n2/d2
    return n1 * d2 == n2 * d1
\end{lstlisting}
\end{minipage}%
    \hfill
\begin{minipage}[t]{\linewidth}
    \textbf{\textsc{CDDRefactorER} - Refactored using \texttt{gpt-5-nano}}:
    
\begin{lstlisting}
def parallel_lines(line1, line2):
    return line1[0]/line1[1] == line2[0]/line2[1]
\end{lstlisting}
\end{minipage}
\end{shaded}

\caption{Original code (top), erroneous baseline refactoring (middle), and correct \textsc{CDDRefactorER} refactoring (bottom).}

\label{fig:cdd_example1}
\end{figure}

\begin{figure}[tb]
\definecolor{boxcolor}{HTML}{CEBD99}
\colorlet{shadecolor}{boxcolor}
\begin{shaded}
\begin{minipage}[t]{\linewidth}
\textbf{Original Code:}
\begin{lstlisting}
def nth_even(n):
    if n==1:
        return 0
    if n==2:
        return 2
    if n==3:
        return 4
    else:
        return n*2-2
\end{lstlisting}
\end{minipage}%
\hfill

\begin{minipage}[t]{\linewidth}
\textbf{Baseline - Refactored using \texttt{gpt-5-nano}:}
\begin{lstlisting}
def nth_even(n):
    """
    Return the n-th even number (1-indexed).
    Examples: 1 -> 0, 2 -> 2, 3 -> 4
    """
    if n < 1:
        raise ValueError("n must be a positive number")
    return (n - 1) * 2
            \end{lstlisting}
    \end{minipage}%
    \hfill
    \begin{minipage}[t]{\linewidth}
\textbf{\textsc{CDDRefactorER} - Refactored using \texttt{gpt-5-nano}:}

\begin{lstlisting}
def nth_even(n):
    return (n - 1) * 2
\end{lstlisting}
\end{minipage}
\end{shaded}

\caption{Original code (CC=4, CogC=4) (top), baseline refactoring (CC=2, CogC=1) (middle), and \textsc{CDDRefactorER} refactoring with lowest complexity (CC=1, CogC=0) (bottom).}

\label{fig:cdd_example2}
\end{figure}

\subsubsection{\textbf{Illustrative Examples}}
To complement the quantitative analysis, we present two representative examples that highlight qualitative differences between baseline and \textsc{CDDRefactorER}. 

Figure~\ref{fig:cdd_example1} shows a function that checks whether two lines are parallel. Under unconstrained prompting, refactoring introduces additional logic based on inferred domain assumptions, altering program behavior and causing test failures. In contrast, \textsc{CDDRefactorER} preserves the original implementation.

Figure~\ref{fig:cdd_example2} presents a case where both approaches preserve functional correctness. The original implementation computes the n-th even number using multiple conditional branches for specific values of n resulting in unnecessary control-flow complexity. The baseline refactoring improves the implementation by introducing a direct mathematical formula and adding input validation. In contrast, \textsc{CDDRefactorER} further simplifies the code by expressing the same formula in its minimal form, removing additional checks and producing the lowest complexity among the three versions.

\smallskip
\noindent\fbox{
\parbox{0.46\textwidth}{
        \textit{\underline{Summary of RQ2.}} \textsc{CDDRefactorER} consistently produces safer and more stable refactorings than unconstrained prompting. Across datasets and models, it significantly reduces refactoring failures, limits increases in cognitive and cyclomatic complexity, and preserves greater structural similarity to the original code. These results indicate that CDD principles and the imposed constraints enable safer and controlled automated refactoring.
    }
}

\section{Human Study}
\textbf{RQ3: \textit{How does systematic automatic refactoring using \textsc{CDDRefactorER} affect novice programmers’ ability to understand code?}}

RQ2 established that \textsc{CDDRefactorER} produces structurally more controlled refactorings than unconstrained prompting --- reducing complexity-increasing transformations and preserving greater structural similarity to the original code. RQ3 examines whether these structural properties translate into measurable differences in novice comprehension. Specifically, lower cyclomatic and cognitive complexity are hypothesized to reduce the control-flow reasoning burden on novices, while higher CodeBLEU similarity is hypothesized to preserve structural familiarity, together supporting comprehension~\cite{campbell2018cognitive, munoz2020empirical, wiese2019linking}. We conducted a controlled between-subjects study with 20 first-semester computer science students as described in Section~\ref{sec:human-study} to test this.

\noindent
\subsubsection*{\textbf{Findings.}}
Table~\ref{tab:comprehension} summarizes the average code comprehension ratings before and after exposure to CDD-refactored code. Across all four measured dimensions, participants reported higher comprehension after reviewing the refactored versions. The largest improvement was observed in Function Identification, which increased from 2.97 to 3.90 (+31.31\%), indicating that refactoring substantially aided participants in recognizing functional roles within the code. Ratings for Code Structure for Readability also improved notably, rising from 3.17 to 3.87 (+22.0\%), suggesting clearer structural organization. More moderate but consistent gains were observed for Purpose Understanding, which increased from 3.23 to 3.80 (+17.65\%), and Logic Flow Comprehension, which improved from 2.93 to 3.50 (+19.45\%). Overall, these survey feedback suggest that cognitively guided refactoring enhances novice programmers’ perceived understanding of code, particularly in terms of functional decomposition and structural clarity.

\noindent
\subsubsection*{\textbf{Qualitative Feedback.}}
Open-ended post-test responses consistently indicated that \textsc{CDDRefactorER} improved novice programmers’ perceived clarity and organization of code. Participants frequently attributed these improvements to clearer structural decomposition, stepwise logic, and more informative naming. For example, one student noted, 
\textit{“[...] Clear names and structure make the logic easy to follow [...].”} (P09). Several participants emphasized that renaming and organization directly supported readability, reporting that 
\textit{“The structured way and meaningful name make the code more easier to read and understand.”} (P15). 
Participants also highlighted the value of explanations and examples accompanying the refactored code. Many described the refactored solutions as more understandable; for example, P05 reported that the solutions were \textit{“easy to understand with example and explanations”} and that \textit{“the explanation with example is great.”}. These responses suggest that combining structural refactoring with contextual explanations further supports comprehension beyond code-level changes alone.

For more advanced problems requiring specialized or less familiar programming concepts, responses revealed both improvement and remaining challenges. In one specific advanced task, during the pre-test, both participants (P02 and P04) reported confusion when interpreting compact or non-obvious expressions, noting that certain conditions and operations were difficult to reason about. For the same problems, in the post-test, one participant indicated substantial improvement, stating, \textit{“I was not understanding the code earlier. But now I understood the code fully.”} (P19). However, not all difficulties were resolved as the other participant continued to report challenges even after refactoring, explaining, \textit{“i don't know why but i cannot understand the part of while loop [...] may be my concept is not clear.”} (P05). These responses suggest that, while refactoring can alleviate structural and readability issues, it cannot fully resolve gaps in the learners' understanding of advanced programming concepts they may not be familiar with.

\begin{table}[tb]
\centering
\caption{Code comprehension ratings on human study.}
\label{tab:comprehension}
\setlength{\tabcolsep}{1pt} 
\small
\begin{tabular}{|l|c|c|c|}
\hline
{5-Point Likert Scale Questions} & {Before} & {After} & {Change} \\
\hline
Function Identification & 2.97 & 3.9 & +31.31\% \\
Code Structure for Readability & 3.17 & 3.87 & +22.0\% \\ 
Purpose Understanding & 3.23 & 3.8 & +17.65\% \\
Logic Flow Comprehension & 2.93 & 3.5 & +19.45\% \\
\hline
\end{tabular}
\end{table}

\smallskip
\noindent\fbox{
\parbox{0.46\textwidth}{
        \textit{\underline{Summary of RQ3.}} Results from the human study show that novices report higher code comprehension after interacting with \textsc{CDDRefactorER}, with notable improvements in function identification, structural readability, and understanding of program logic. These findings suggest that cognitively guided refactoring can support novice comprehension by reducing cognitive overload.
    }
}

\section{Implications}

Our study demonstrates that cognitively guided automated refactoring can meaningfully support novice code comprehension when structural changes are constrained by cognitive principles. The findings have implications for educational practice, tool design, and future research.

\subsubsection*{\textbf{Implications for Educational Practice}}
Results from the human study indicate that cognitively guided refactoring yields the largest comprehension gains in function identification and structural readability. This suggests that refactoring can act as an effective instructional scaffold for helping novices recognize functional decomposition and navigate control-flow, two areas that are consistently reported as challenging for early learners~\cite{busjahn2011analysis, peitek2021program, sellitto2022toward, siegmund2017measuring}.
However, qualitative feedback shows that refactoring alone does not resolve gaps in conceptual understanding, particularly for unfamiliar programming constructs~\cite{wiese2019linking}. Consequently, automated refactoring should complement, rather than replace, foundational instruction~\cite{berssanette2021cognitive, duran2022cognitive}. We recommend integrating refactoring tools after an initial manual comprehension phase, where students first attempt to understand code independently~\cite{carneiro2024investigating}. This sequencing encourages active reasoning while allowing refactored code to serve as a confirmatory or corrective artifact rather than a primary source of understanding~\cite{macneil2023experiences}.

We recommend a three-step classroom workflow: (1) students first attempt to understand the original code independently, surfacing genuine points of confusion~\cite{sweller1988cognitive, prather2023weird}; (2) instructors use \textsc{CDDRefactorER} to refactor units where confusion is widespread~\cite{Hasan2026LearningProgramming}; and (3) refactored and original versions are reviewed side-by-side, with explicit discussion of structural changes to avoid over-reliance on generated outputs~\cite{wiese2019linking, 
carneiro2024investigating, prather2023weird}.

\subsubsection*{\textbf{Implications for Tool Design}} Across datasets and models, \textsc{CDDRefactorER} substantially reduces refactoring failures and limits structural regressions compared to unconstrained prompting~\cite{alomar2023automating, piao2025refactoring}. These results indicate that cognitive principles should be treated as first-class design constraints in refactoring tools intended for novices~\cite{souza2020toward, pinto2023cognitive, pinto2021cognitive}.

The observed increase in structural similarity, as measured by 
CodeBLEU~\cite{ren2020codebleu}, suggests that novice-oriented tools should prioritize incremental and localized refactorings over aggressive restructuring~\cite{wiese2019linking, hermans2016code}. In particular, refactoring strategies that extract well-named helper functions and reduce unnecessary nesting appear especially effective~\cite{scalabrino2016improving, sellitto2022toward}, given the significant improvement in function identification reported by participants. Tool designers should therefore emphasize bounded transformations that improve clarity while preserving familiarity with the original code structure~\cite{wiese2019linking, souza2020toward}.

\subsubsection*{\textbf{Implications for Researchers}}
The results show that imposing cognitively motivated structural constraints leads to refactoring outcomes that differ systematically from unconstrained prompting in terms of correctness, structural stability, and comprehension-relevant properties~\cite{campbell2018cognitive, munoz2020empirical}. This indicates that cognitive constraints should be treated as explicit experimental factors when studying automated refactoring systems, rather than as implicit design choices~\cite{morales2020repor, alomar2025chatgpt}. Finally, the principles demonstrated here motivate future research on applying cognitively guided constraints to related program transformation tasks, such as code smell detection and remediation~\cite{hermans2016code}, automated program repair~\cite{le2019automated}, code generation and explanation~\cite{macneil2023experiences, roziere2023code}, and other code transformation settings where trade-offs between correctness, structural change, and human understanding are central~\cite{sellitto2022toward, fakhoury2018effect}.

\section{Threats to Validity}

We acknowledge several key threats to validity in our study. Below we describe them:

\subsubsection*{\textbf{Construct Validity}}
Our study uses cyclomatic complexity and cognitive complexity as proxies for structural difficulty and cognitive effort during code comprehension. While researchers widely use these metrics and they are theoretically grounded, these are static measures and do not directly capture human cognitive processes. To mitigate this limitation, we complement metric-based analysis with a controlled human study that directly measures novice comprehension across multiple dimensions. For functional correctness, we rely on the test suites provided by the MBPP and APPS datasets. Although these test suites may not exhaustively cover all edge cases, they provide a consistent and widely accepted basis for evaluating behavior preservation across both baseline and CDD-guided refactoring settings.

\subsubsection*{\textbf{Internal Validity}}
Internal validity concerns whether the observed differences in outcomes can be attributed to the refactoring approach rather than to confounding factors. Because the human study uses separate pre-test and post-test groups composed of different participants, individual differences in prior knowledge and programming ability as well as cognitive capacity (e.g., working memory and ability to manage complex control-flow) may influence comprehension outcomes independently of the refactoring condition. Although participants were drawn from the same course level and task difficulty was balanced across groups, such differences in knowledge and mental capacity cannot be fully controlled and may pose a threat to internal validity. In addition, variability in how participants interacted with \texttt{CDDRefactorER}, including differences in how refactored code was examined or interpreted, may affect comprehension results. Identical procedures and system settings were used to reduce procedural bias and limit systematic differences between conditions.

\subsubsection*{\textbf{External Validity}}
The findings of this study are grounded in novice-level programming tasks drawn from the MBPP and APPS datasets, which primarily consist of small, self-contained algorithmic problems. While these tasks are appropriate for studying novice code comprehension, they may not fully represent the complexity of real-world software systems involving larger codebases, multiple files, or domain-specific frameworks. The human study focuses on first-year undergraduate students, which matches the intended target population but limits generalization to more experienced programmers. In addition, results are based on two language models and a specific refactoring configuration, and outcomes may differ with other models, programming languages, or instructional contexts.

\subsubsection*{\textbf{Conclusion Validity}}
The human study involves a relatively small number of participants, which limits statistical power and the ability to detect subtle effects. However, the observed improvements are consistent across multiple comprehension dimensions and are supported by qualitative feedback, increasing confidence in the reported trends. We conducted statistical analyses using appropriate non-parametric tests, and reported effect sizes to support interpretation beyond significance testing alone.

\section{Conclusion and Future Work}
This work shows that cognitively guided automated refactoring improves both refactoring safety and novice code comprehension compared to unconstrained prompting. Across MBPP and APPS, \textsc{CDDRefactorER} reduced refactoring failures by 54.40-71.23\% and substantially lowered the rate of structural regressions. On APPS, cognitive complexity increases fell from 29.08\% to 12.32\% for \texttt{gpt-5-nano} and from 21.8\% to 10.12\% for \texttt{kimi-k2}, while cyclomatic complexity increases dropped from 26.56\% to 6.18\% and from 15.0\% to 3.66\%, respectively. \textsc{CDDRefactorER} also preserved greater structural similarity, with median CodeBLEU scores rising by 75.6–122.7\%, reflecting more controlled and stable transformations.

In the human study, cognitively guided refactoring led to higher self-reported comprehension across all dimensions and reduced cognitive overload that had arisen from code understanding, with the largest gains in function identification (+31.31\%) and structural readability (+22.0\%), followed by improvements in logic flow (+19.45\%) and purpose understanding (+17.65\%). This indicates that constraining refactoring with cognitive principles improves comprehension-relevant structure without sacrificing correctness.

Future work should evaluate whether these gains persist in longitudinal settings, scale to larger multi-file codebases, and generalize to other program transformation tasks such as program repair, code translation, and educational code generation. Further studies should also examine how varying cognitive thresholds affects the trade-off between simplification and structural familiarity.

\balance
\bibliographystyle{ACM-Reference-Format}
\bibliography{references}
\end{document}